# Grain Boundary Segregation Predicted by Quantum-Accurate Segregation Spectra but not by Classical Models

Malik Wagih[1*†], Yannick Naunheim[1*], Tianjiao Lei[1*], and Christopher A. Schuh[1,2‡]

[1] Department of Materials Science and Engineering, Massachusetts Institute of Technology,
77 Massachusetts Avenue, Cambridge, MA 02139, USA

[2] Department of Materials Science and Engineering, Northwestern University,
2145 Sheridan Road, Evanston, IL 60208, USA

**Abstract**

In alloys, solute segregation at grain boundaries is classically attributed to three driving forces: a high solution enthalpy, a high size mismatch, and a high difference in interfacial energy. These effects are generally cast into a single scalar segregation energy and used to predict grain boundary solute enrichment or depletion. This approach neglects the physics of segregation at many competing grain boundary sites, and can also miss electronic effects that are energetically significant to the problem. In this paper, we demonstrate that such driving forces cannot explain, nor thus predict, segregation in some alloys. Using quantum-accurate segregation spectra that have recently become available for some polycrystalline alloys, we predict strong segregation for gold in aluminum, a solvent-solute combination that does not conform to classical driving forces. Our experiments confirm these predictions and reveal gold enrichment at grain boundaries that is two orders of magnitude over the bulk lattice solute concentration.

## 1. Introduction

The segregation of solute atoms at grain boundaries (GBs) impacts many material properties both structural and functional [1]. This is true even for dilute solute additions, since GB segregation elevates the local concentration of solutes at the GB, sometimes by orders of magnitude relative to the bulk [2–10]. As a result, GB segregation has become an important alloy design tool [11–13]. However, the effectiveness of such a tool requires that it accurately screen for and quantify segregation across the alloy space.

There are many ways to compute the segregation energy when a solute atom is placed at a GB site vis-à-vis the bulk [14–22]. For polycrystals, or for alloy design, there is a range of GB sites, and there are two main approaches to predicting the resulting level of GB segregation that one would expect to see overall: (i) the classical McLean-style approach [23–25], and (ii) the more recent and data-rich spectral approach [26–30]. In the classical approach, the GB network is treated as a single unit with a single site-type, and the energetic preference for a solute atom to segregate at that site is described using a single average segregation enthalpy, $\Delta \overline{H}^{seg}$. At higher concentrations that energy is sometimes corrected to account for solute-solute interactions [31,32], but the single site-type assumption remains the defining feature of models of this type. The simplicity of this approach is a significant advantage: computing $\Delta \overline{H}^{seg}$ is all that is needed to screen for segregation in binary alloys in the McLean framework. A number of approximations for calculating this value have been proposed in the literature, with Murdoch and Schuh [25] providing the largest database of which we are aware; using a Miedema-based [33–35] approach, they calculated $\Delta \overline{H}^{seg}$ for ~2500 alloys, providing a comprehensive toolbox for screening for GB segregation. Nevertheless, the classical approach is a major oversimplification of the true disordered nature of the GBs [36], since it omits the fact that GBs are characterized by a wide variety of local atomic environments that can attract or repulse solute atoms to different degrees.

---

[*] These authors contributed equally to this work.
[†] Current address: Lawrence Livermore National Laboratory, 7000 East Avenue, Livermore, CA 94550, USA.
[‡] Corresponding author. Email address: schuh@northwestern.edu .

The more recent spectral approach, on the other hand, accounts for sitewise variation by treating a spectrum of segregation energies (enthalpies) for the collection of GBs in a material. In this representation, the effort to characterize segregation in a binary alloy becomes more complex, since we need to compute the full spectrum of segregation energies in a polycrystal, rather than a single average value. Our group has lately used machine learning algorithms in combination with advanced atomistic simulation methods to compute the spectrum and a variety of segregation thermodynamic quantities in polycrystals [28–30,37–41]. The resulting large databases [29,30] cover many hundreds of binary alloys and provide a richer view of the range of segregation states in binary polycrystals. Some of these data are even computed to the level of quantum-accuracy [30]. And, with the rapid advances in machine learning tools for materials science [42–46], including their applications to the segregation problem [19,29,30,47–51], we expect such databases to continue to improve in both accuracy and chemical complexity. In fact, the spectral method of modeling GB segregation has now been developed to the point where it can guide alloy screening efforts [52]. However, the spectral approach to GB segregation has so far remained a theoretical and computational effort; it is not yet connected to experiments. As such, the purported advantages of the method vis-à-vis the classical McLean approach remain untested.

A great deal of simulation work has shown that the classical McLean model of GB segregation fails to capture the correct physics of segregation [19,28,37,53,54], as it fails to account for the correct enthalpic and entropic contributions to segregation. Using a single, average segregation energy fitted at one set of conditions can lead to extremely large extrapolation errors when the model is used at any other composition [37,39], temperature [28,40,54], or grain size [55]. In fact, the functional form of the McLean model simply cannot fit data across a wide range of composition, temperature and grain size space. The spectral model is therefore expected to provide substantially improved quantitative power by comparison, and in fact for some alloys it may make distinctly different predictions than the classical model. This in turn can provide test cases to demonstrate the added value of the spectral approach, which is important at the moment because the spectral model has been developed theoretically and computationally, but not yet directly compared with any targeted experiments that validate its power.

The purpose of the present paper is to address this gap by design: we specifically identify an alloy – Al(Au) – in which the classical model and the spectral model predict widely disparate GB segregation behavior, which permits an experimental test as to which model's predictions are more closely followed. We focus, by design, on dilute binary alloys, both experimentally and computationally, to eliminate any additional complexities that may arise from solute-solute interactions beyond the dilute-limit or interactions among multiple competing solutes (as in multinary alloys). By doing so, we focus on contrasting the difference between (i) the assumption of a single site-type at the GBs as in McLean-style models, and (ii) the multiple site-types treatment of the spectral models. This approach not only provides a first validation of quantum-accurate segregation spectra [30], it also points to critical physics missed by classical modeling and therefore opens a pathway to more accurate defect thermodynamics in alloys more broadly.

## 2. Au segregation in Al: classical versus spectral approaches

In Ref. [30], we presented a database of quantum-accurate[1] segregation enthalpy spectra for ~40 solutes in fcc Al. These are the most accurate computed GB segregation data to date for polycrystals, and the best set from which to identify an ideal experimental test case. Among the ~40 alloys in that set, we identify Al(Au) as a model case, as we develop in more detail in what follows. This alloy, to the best of our knowledge, has not been previously examined for GB segregation.

We turn our attention first to classical McLean-type one-parameter predictions for Au segregation in Al, before comparing it to the spectral findings from Ref. [30]. In his early treatment of GB segregation, McLean [23] assumed that the driving force for segregation comes from relieving the elastic strain energy introduced by the solute atom in the lattice due to its atomic size mismatch, which is expressed as follows for an A(B) binary alloy [56]:

$$\Delta H_{B\,in\,A}^{seg\,[Mclean]} = -\Delta H_{B\,in\,A}^{elastic} = -\frac{24\pi K_A G_B r_B r_A (r_B - r_A)^2}{3K_A r_A + 4G_B r_B} \quad (1)$$

where K is the bulk modulus (typically taken from the small-strain linear response to hydrostatic stresses in the standard crystal environment of each species), G is the shear modulus, and r is the atomic radius. We note that, in the notation of this paper, a negative segregation energy indicates an energetic preference for the solute (B) to segregate to GBs in the solvent (A).

Since the squared difference in atomic radii is always positive, $(r_B - r_A)^2$, Eq. (1) predicts all solutes to segregate to different degrees. Over the years, there have been multiple efforts to rectify this simplification, with the common theme for such models being to account for, in addition to the elastic component, an interfacial component that comes from the difference in pure (unalloyed) GB energies for the solvent and solute, and a chemical component that comes from the enthalpy of chemical interactions (mixing) between the solvent and solute. The model by Murdoch and Schuh [25], inspired by earlier work by Miedema for surface segregation [34,35], accounts for all three contributions to GB segregation as follows:

$$\Delta H_{B\,in\,A}^{seg} = c\left[-\Delta H_{B\,in\,A}^{sol} + c_0 \gamma_B^S V_B^{2/3} - c_0 \gamma_A^S V_A^{2/3}\right] - \Delta H_{B\,in\,A}^{elastic} \quad (2)$$

where $\Delta H_{B\,in\,A}^{sol}$ is the dilute-limit enthalpy of solution (without the elastic mismatch, i.e., the chemical interactions component, which is equivalent to the enthalpy of solution in the liquid) of metal B dissolved in metal A as defined by Miedema [34]; c is a constant equal to 0.12 [25]; $c_0 \gamma^S V^{2/3}$ is the surface enthalpy of a pure (unalloyed) metal, also as defined by Miedema [34], where $c_0$ is a semi-empirical constant, $\gamma^S$ is the surface energy of the pure metal, V is the atomic volume, and $\Delta H_{B\,in\,A}^{elastic}$ is the elastic lattice strain energy as defined in Eq. (1). In Table 1, we list the parameters used to evaluate Eq. (2) for Au in Al. We arrive at a value of $\Delta H_{Au\,in\,Al}^{seg} \approx 20$ kJ/mol; the chemical, interfacial, and elastic terms contribute +12.1 kJ/mol, +8.1 kJ/mol, –0.06 kJ/mol, respectively. This denotes a rather large energetic preference for Au in the lattice, not the GBs; Au is not expected to segregate at all at the GB in Al, according to the classical approach. The observation that classical models based on a single parameter segregation energy, occasionally predict "anti-segregation" such as in this case, is one obvious physical problem with the approach. In a polycrystal with many unique sites, one expects that there would inevitably be some preferential sites that would accept a segregant.

---

[1] In this work, "quantum-accurate" refers to calculations performed at the level of density functional theory (DFT), using the Perdew–Burke–Ernzerhof (PBE) [88] generalized gradient approximation (GGA) for the exchange-correlation functional. We refer the reader to Ref. [30] for more details on computing the segregation spectra.

In Fig. 1(a), we show the learned spectrum of dilute-limit segregation energies for a Au solute at all GB sites in a 20x20x20 nm³ Al polycrystal with an average grain size of ~10 nm and GB site volume fraction of 17%. We refer the reader to Ref. [30] for more details on the computational procedure, but a key point here is that the data are learned from quantum-accurate simulations of the most important statistically representative sites in a true polycrystalline environment. The use of a multiscale quantum mechanical / molecular mechanical (QM/MM) approach [57–59] permits the addressing of complex polycrystalline sites, while the sites selected are based on detailed analysis of millions of sites and permit an accurate estimation of the full spectrum at reasonable computational cost. The spectrum in Fig. 1(a) was published previously in Ref. [30], and here we will explore its features and predictions more closely.

Fig. 1(a) shows that 97% of GB sites in a randomly oriented Al polycrystal are expected to be enthalpically favorable to Au segregation, with 25% of sites having a segregation energy of less than –22 kJ/mol, which for preliminary screening purposes signifies a strongly segregating system. This is markedly different, in fact, fundamentally opposite, segregation behavior as compared with the classical prediction.

Table 1: Parameters used to evaluate $\Delta H^{seg}_{B\ in\ A}$ using Eq. (2)

|  | Al | Au |
|---|---|---|
| K [GPa] [60] | 72.2 | 173.2 |
| G [GPa] [60] | 26.6 | 27.6 |
| r [Å] [60] | 1.432 | 1.442 |
| $c_0 \gamma^S V^{2/3}$ [kJ/mol] [34] | 220.5 | 288.0 |
| $\Delta H^{sol}_{Au\ in\ Al}$ [kJ/mol] [34] | — | –100.8 |

To quantify the disagreement between the two approaches, we compute the predicted enrichment at the GBs for a polycrystal of 100 nm grain size at a temperature of 300 °C. For the classical approach, the equilibrium average GB solute concentration ($\overline{X}^{gb}$) is obtained by solving the following McLean-style isotherm [23]:

$$\overline{X}^{gb} = \left[1 + \frac{1-X^{bulk}}{X^{bulk}} \exp\left(\frac{\Delta H^{seg}_{B\ in\ A}}{k_B T}\right)\right]^{-1} \quad (3)$$

where $X^{bulk}$ is the bulk (intra-grain) solute concentration; $k_B$ is Boltzmann's constant; and T is the temperature. We note that in this paper, we use "H" to denote both enthalpies and internal energies (which can be computed directly from atomistic simulations) since both are equivalent in solids [61]. For the spectral model, the equilibrium segregation state is obtained from [26–28,62,63]:

$$\overline{X}^{gb} = \int_{-\infty}^{\infty} F^{gb}_i(\Delta H^{seg}_i) \cdot \left[1 + \frac{1-X^{bulk}}{X^{bulk}} \cdot \exp\left(\frac{\Delta H^{seg}_i}{k_B T}\right)\right]^{-1} d(\Delta H^{seg}_i) \quad (4)$$

which is similar to Eq. (3) but is now expressed in terms of an integral over the distribution function, $F^{gb}_i(\Delta H^{seg}_i)$ that describes the spectrum of segregation energies ($\Delta H^{seg}_i$) found in the polycrystal, with $\Delta H^{seg}_i$ being the segregation energy for GB site (i). $F^{gb}_i(\Delta H^{seg}_i)$ is defined as:

$$F_i^{gb}(\Delta H_i^{seg}) = \frac{1}{\sqrt{2\pi}\,\sigma} \exp\left[-\frac{(\Delta H_i^{seg} - \mu)^2}{2\sigma^2}\right] \text{erfc}\left[-\frac{\alpha(\Delta H_i^{seg} - \mu)}{\sqrt{2}\,\sigma}\right] \quad (5)$$

where $\mu$, $\sigma$, $\alpha$ describe the characteristic energy, the width, and the shape, respectively, for a best-fit skew-normal distribution; for the quantum-accurate spectrum, $F_i^{gb}(\Delta H_i^{seg})$ is shown as a solid blue line in Fig. 1(a). In Fig. 1(b), we show the predicted solute enrichment at the GB, $\overline{X}^{gb}/X^{bulk}$, for the classical and quantum-accurate spectral approaches. Whereas the classical approach predicts depletion of Au at the GB, the spectral approach predicts strong enrichment. This strong disagreement provides an excellent test case for experimental validation of the nominally more-physical quantum-accurate spectral model, to which we now turn our attention.

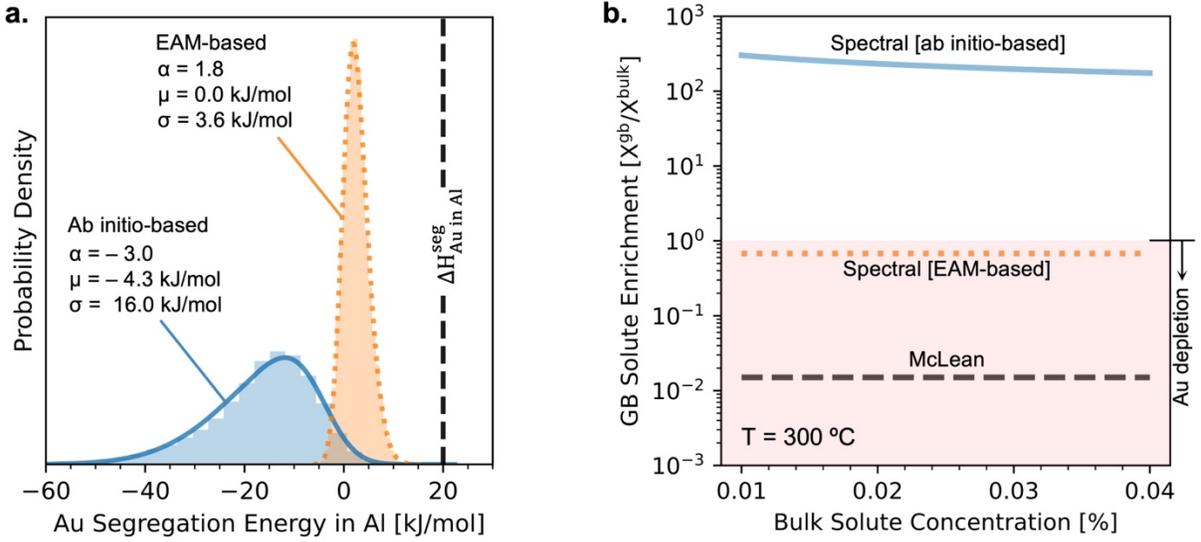

Fig. 1: (a) A comparison of GB solute segregation energies computed for Au in Al computed using the ab initio-based framework from Ref. [30]; the embedded atom (EAM) interatomic potential [64] from Ref. [29]; and the single McLean-style average value obtained from Eq. (2). (b) Predictions of the equilibrium solute enrichment ($\overline{X}^{gb}/X^{bulk}$) at the GBs at 300 °C for the spectral model (Eq. (4)) using both the ab initio-based and the EAM-based spectra, and the classical McLean-style model (Eq. (3)) using $\Delta H_{Au\,in\,Al}^{seg} \approx 20$ kJ/mol.

## 3. Experiments

We fabricated a dilute Al–2 atomic % Au alloy by mechanical alloying [65] using high-energy ball milling of elemental powders (Al 99.9% purity, 44 µm particle size; Au 99.96% purity, 3-5 µm particle size, both from Fisher Scientific). Milling was performed using a SPEX 8000D Mixer/Mill for 20 hours in a glovebox under a high-purity argon environment to minimize oxygen exposure. Hardened steel grinding jar and media were used with a 10:1 ball-to-powder ratio and 0.1 ml/g of dry ethanol as a process control agent.

The as-milled composition of the powders was determined using an energy-dispersive x-ray spectroscopy (EDS) detector performed using a Zeiss Merlin High-resolution scanning electron microscope (SEM). EDS measurements provide a composition of Al–1.88 ± 0.08 at. % Au, which is very close to the nominal composition. The EDS measurements also suggest that oxygen content was kept below about ~4 wt. %. High-energy ball milling produces micron-sized powder particles of solid-solution Al(Au), with an average grain size of ~30 nm as determined by X-ray diffraction (XRD) measurements. XRD was performed using the Panalytical X'Pert Pro X-ray diffractometer using a Cu-K$_\alpha$ source. It is useful for the present evaluation to work with a nanocrystalline structure; it allows us to characterize a significant number of GBs in a small sample volume, and provides short diffusion distances for solute atoms to equilibrate and occupy their preferred sites upon annealing.

Following milling, we thermally annealed a portion of the powder at 300 °C for 1 hour under a reducing atmosphere of Ar–4%H to minimize oxygen exposure. The combination of the elevated temperature (~60 % of the liquidus temperature for the alloy [[66]]) and the short diffusion distance (tens of nanometers) to GBs ensures that, first, the GBs are relaxed from the initial as-milled conditions; second, that the solute distribution in the system can equilibrate – Au has a mean bulk diffusion distance of ~700 nm after 1 hour based on diffusivity measurements of Au in Al from Ref. [67]; and third, secondary phases, if any, can form. The annealed sample was characterized by scanning transmission electron microscopy (STEM) combined with EDS using a Thermo Fisher Scientific Themis Z probe aberration-corrected STEM operated at 200 kV. To prepare TEM specimens, the focused ion beam lift-out method was employed [68], and a final polish of 5 kV and 15 pA was used to minimize damage from the Ga ion beam; these parameters should result in a local temperature increase of only a few degrees, and thus we do not expect the implantation of Ga ions to alter the segregation state of Au, given that Au diffusion in Al is extremely low at room temperature.

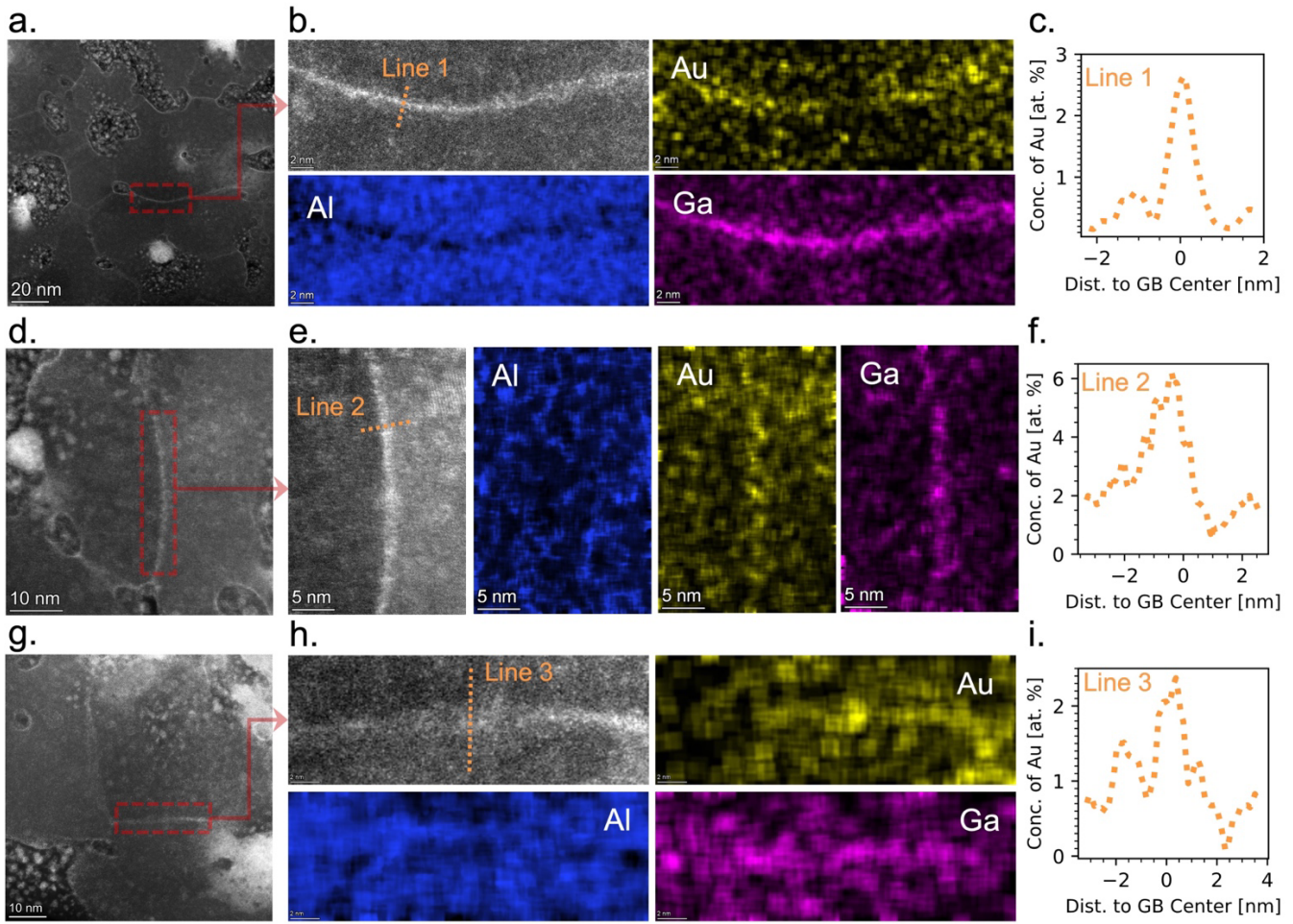

Fig. 2: For three different GBs, we show in (a), (d), and (g), an STEM micrograph of the section of the microstructure from which the GBs are sampled; in (b), (e), (h), a close-up of the GBs along with EDS maps of Al, Au, and Ga; and in (c), (f), and (i), line scans across the GBs that show clear enrichment of Au at the GBs.

In Fig. 2(a), we show a high-angle annular dark-field (HAADF) STEM micrograph of the microstructure for a section of the annealed sample; since Au is a much heavier element than Al, it appears bright in the micrograph. Fig. 2(a) shows a nanocrystalline structure (grain size < 100 nm; an average of ~50 nm) with secondary Au-enriched phases. Al$_2$Au intermetallic compounds are expected equilibrium phases for the Al-Au system [66], and indeed large precipitated second phases richer in gold than the majority Al grains are observed. Only ~0.01 at. % Au is expected to remain in solid solution at 300 °C [66,69]; this is the quantity of gold that should be used in Eqs. (3) and (4) for predictions of the partitioning between bulk and GBs. While EDS cannot resolve such a low concentration in the bulk solid solution, we expect Au concentrations to be extremely low given the known low solubility of the system, the significant equilibration time used here, and the very short diffusion distances to the GBs.

As for Al$_2$Au, because it is a highly stable compound, with an experimentally measured formation energy of –41.4 kJ/mol [66,70], solute atoms are not expected to leave it for the GB, i.e., the intermetallic compounds in the system are not expected to contribute to solute enrichment, if any, at the GB. To roughly quantify this: the energetic benefit of a solute atom segregating from the compound to a GB site (i) is $\approx \Delta H_i^{seg} - \left[\frac{1}{x_s^c} \Delta H_{Al_2Au}^{form} - \Delta H_{Au\ in\ Al}^{sol+elastic}\right]$, where $x_s^c$ is the compound stoichiometry (1/3 in this case), $\Delta H_{Al_2Au}^{form}$ is the compound formation energy, and $\Delta H_{Au\ in\ Al}^{sol+elastic}$ is the dilute solution energy (both chemical and elastic components) for Au in Al. A simple $\exp[-\Delta H/(k_B T)]$ analysis using all GB sites leads to – at maximum – a Au solute concentration at the GB of $\bar{X}^{gb}(\text{from Al}_2\text{Au}) \approx \int_{-\infty}^{\infty} \left[F_i^{gb}(\Delta H_i^{seg}) \cdot \exp\left[-\left(\Delta H_i^{seg} - (\Delta H_{Al_2Au}^{form}/x_s^c - \Delta H_{Au\ in\ Al}^{sol+elastic})\right)/(k_B T)\right]\right] d(\Delta H_i^{seg}) \approx$ 0.03% [2], which is a negligible contribution to the observed Au solute concentration at the GB as we show next.

Fig. 2(b) shows a close-up of a single GB from Fig. 2(a), along with an EDS map of Al, Au, and Ga. Ga is a known strong segregator to interfaces and disordered regions in Al [71,72], and is therefore a fiducial marker of GBs. Fig. 2(b) shows clear enrichment of Au at the GB, which is quantified in Fig. 2(c) with an EDS line scan across the GB (dashed line in Fig. 2(b)) that shows over two orders of magnitude enrichment of Au at the GB over the bulk lattice concentration of 0.01 at. % Au. We further corroborate this observation in Fig. 2(d-f) and (g-h) with EDS maps and line scans of two additional, typical GBs sampled from different regions of the microstructure, both of which exhibit strong enrichment of Au.

To further quantify the segregation of Au, we took 120 line scans across 11 different GBs, with the distance between two adjacent lines being at least 1–2 nm. Each of the 11 GBs was inspected to ensure that it was a boundary between two Al solid solution grains, without any oxides in the analysis area. For each line scan, an integration width of 4 pixels (=0.2 nm) was used. Subsequently, the Au concentration at each GB site was obtained by taking the average of those data points corresponding to the GB location. Fig. 3 shows the cumulative probability distribution of Au concentrations observed in all 11 GBs.

---

[2] To make the calculation self-consistent, all parameters are evaluated from first-principles. $\Delta H_{Al_2Au}^{form}$ = –47.8 kJ/mol is obtained from Ref. [89]. To compute $\Delta H_{Au\ in\ Al}^{sol+elastic}$, we performed non spin-polarized DFT calculations using the GGA-PBE [88] exchange-correlation functional, using the code GPAW [90,91]. A plane wave basis set is used, with an energy cutoff of 400 eV. The Brillouin zone is integrated with a Monkhorst-Pack grid using a k-point density of 3 points per Å$^{-1}$, and a Marzari-Vanderbilt smearing [92] of 0.2 eV. $\Delta H_{Au\ in\ Al}^{sol}$ is defined as $H(Al_{N-1}Au) - [(N-1) \cdot H_{Al}^{eq} + H_{Au}^{eq}]$, where $H(Al_{N-1}Au)$ is the relaxed total energy of an N-atom Al supercell with one solute Au atom replacing an Al atom, N is 256 atoms (4x4x4 repeats of the fcc cubic cell for Al), and $H_{Al}^{eq}$, $H_{Au}^{eq}$ are the energies per atom for both Al and Au, respectively, in their unalloyed perfect supercells. Structural relaxations are performed [93,94] using the FIRE [95] algorithm using a force convergence criterion of 0.01 eV/Å. We obtain $\Delta H_{Au\ in\ Al}^{sol+elastic}$ = –70.2 kJ/mol.

There are several key findings in Fig. 3. First, all 11 GBs show clear enrichment in Au. Although the accuracy of EDS is not high enough to resolve concentrations below the bulk level of ~0.01% and thus would not be expected to reveal local depletion of Au, the direct measurement of enrichment at all sites sampled is strong collective evidence for segregation, and is contrary to the depletion expected by the classical model in reference to Fig. 1. Second, the data in Fig. 3 reflect the richness of the segregation spectrum in two complementary ways. Most obviously, the 11 GBs sampled here show a wide variety of segregation behavior, with median values that range from ~0.5 to ~3 at.%. Each GB is different than the next in terms of its ability to segregate Au, by as much as an order of magnitude in concentration. However, for any individual GB there is a second source of spectrality at the site level; every boundary has a wide spread of Au concentrations which are often around 1 at.% in width. The EDS pixel size is ~0.2 nm and the measurement integrates over depth through the sample thickness, so this variation is not attributable to the atomic site level directly. It is an interesting direction for future work to assess how local measurements like these could be used to compare experiments more directly with computed site-wise spectra such as that in Fig. 1(a). However, the fact that GB segregation is spectral even within each individual GB is an important qualitative validation for the use of a spectral modeling approach rather than a single segregation energy.

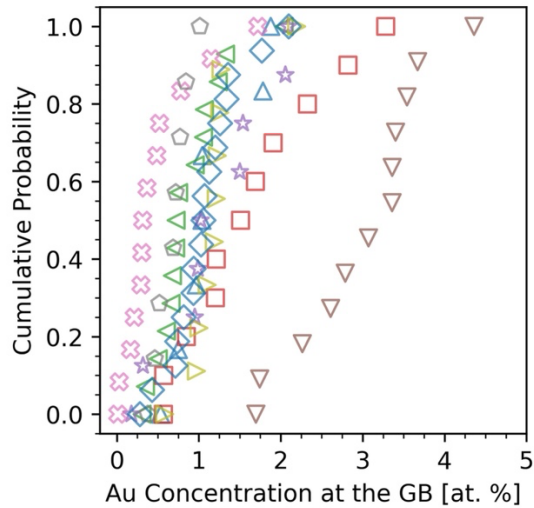

Fig. 3: The cumulative probability distributions of measured Au concentrations using multiple STEM-EDS line scans for 11 different GBs in Al.

## 4. Discussion

As illustrated in reference to Fig. 1(b), the quantum-accurate spectral model correctly predicts the experimentally observed strong Au enrichment at the GBs whereas the McLean approach is incorrect even to the level of predicting the sign of the segregation. To better illustrate this, we combine all 120 line scans from Fig. 3 into a single distribution, and recast it in Fig. 4 in terms of Au enrichment at the GBs, $\bar{X}^{gb}/X^{bulk}$, along with predictions of the spectral and classical models; to calculate enrichment, we take $X^{bulk} = 0.01$ at. % Au (which is the solubility limit for Au in the bulk lattice solid solution). Fig. 4 shows that GB regions are on average enriched by over two orders of magnitude (~$10^2$) with $\overline{X}^{gb}$ = 1.2 at. % Au, and over 10% of the GB locations enriched by >200 times. In comparison, the quantum-accurate spectral model correctly predicts two orders of magnitude (~$3\times10^2$) enrichment of Au at the GB, with $\overline{X}^{gb}$ = 3.0 at. %, which is within the experimental range of Au concentrations, despite being slightly on the higher end. It is thus notable how closely the quantum-accurate spectral model predictions match the experimental results.

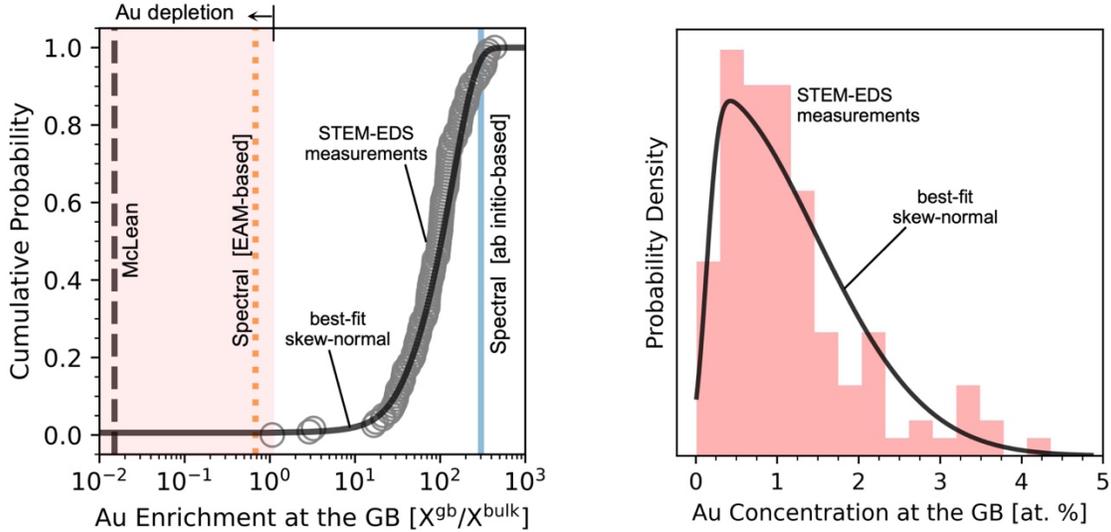

Fig. 4: (a) We collapse all 120 STEM-EDS line scans from the 11 GBs presented separately in Fig. 3 into a single distribution, and show the observed Au enrichment at the GB; we compare the experimental findings (open circles) against predictions of the spectral and classical models, obtained from Eqs. (4) and (3), respectively, using a bulk solute concentration of 0.01 at. % Au. (b) The distribution of measured Au concentrations for all 120 STEM-EDS line scans.

The failure of the classical approach is especially noteworthy here, since all three of the driving forces classically believed to drive GB segregation fail to predict Au segregation in Al. Specifically, the three terms in Eq. (2) reflect driving forces from the reduction of interfacial, chemical, and elastic mismatch energies in the alloy system. We consider each of these in the present case in turn:

- Interfacial energies. Solute segregation is viewed as being promoted by lower solute interfacial energy, i.e., when the solute has a lower GB energy in its unalloyed state as compared with the solvent, then swapping out GB sites from solvent to solute is energy lowering to the GB itself, independently of the elastic and chemical terms. Contrary to this expectation in the present case, Au has stronger intrinsic bonding, a higher melting point, and a higher surface energy. By extension, Au has a slightly higher GB energy as well of 0.37–0.45 J/m$^2$ [73,74] as compared with Al which is 0.30–0.38 J/m$^2$ [75,76]. Miedema's model estimates 0.52 J/m$^2$ for Au and 0.4 J/m$^2$ for Al (if we assume the GB energy to be one-third of the surface energy calculated by Miedema [77]). All of these values speak against Au segregation in Al GBs. One possible exception is provided by density functional theory calculations [78] that predict slightly lower GB energies for specific boundaries in Al as compared to Au; for example, the computed energy for $\Sigma7(3\overline{2}\overline{1})$ is 0.44 J/m$^2$ for Au versus 0.5 J/m$^2$ for Al. However, this difference in GB energies is too slight to affect the situation: it predicts a shift in $\Delta H_{B\ in\ A}^{seg}$ by less than –1 kJ/mol, which is insignificant in Fig. 1(a).

- Solution enthalpy. A higher enthalpy of chemical interactions from solvent-solute mixing generally favors GB segregation. If an A(B) alloy has a positive mixing solution enthalpy, A–B bonds come with an energetic penalty as it increases the system energy. Since GBs are generally less coordinated, i.e., have more broken bonds, the number of A–B bonds (and their energetic penalty) decreases when a B (solute) atom segregates from the bulk lattice to the GB. Thus, the higher the solution enthalpy, the more segregation is expected. The opposite is true for alloys with a negative solution enthalpy since A–B bonds reduce the system energy. Au has a negative mixing enthalpy (which has a similar sign but a different magnitude than the solution enthalpy) with Al, which according to phase diagram data [66,79] is on the order –20 kJ/mol for an equimolar liquid. Computing $\Delta H_{Au\ in\ Al}^{sol}$ from first-principles, we obtain[3] –71 kJ/mol; the Miedema estimated value of –100.8 kJ/mol is not far from it. These values strongly favor solute occupation in the bulk lattice over the less coordinated GB environment, at a level of 12.1 kJ/mol based on Miedema's enthalpies, and 8.5 kJ/mol based on first-principles; a shift of –3.6 kJ/mol in the value $\Delta H_{B\ in\ A}^{seg}$, which is not significant enough to alter its predictions of strong anti-segregation.

- Atomic size mismatch. As in the classical McLean view, GBs are viewed as elastic relaxation sites for solutes of mismatched size. A misfitting solute atom introduces an elastic distortion around it in the rigid ordered bulk lattice. And, since GBs are naturally disordered and have more free volume, they can better accommodate a misfitting solute atom with less or no elastic distortion. In fact, it is generally assumed that the bulk elastic strain energy is completely eliminated when the solute atom replaces a solvent atom at the GB. The larger the misfit, the stronger the segregation. In the case of Al(Au), however, the size mismatch between Au and Al is too small to motivate segregation; an evaluation of Eq. (1) gives $\Delta H_{B\ in\ A}^{elastic} = 0.06$ kJ/mol, i.e., almost zero. Similarly, a first-principles analysis (cf. footnote 3) of local atomic strains introduced by Au in a bulk Al matrix amounts to a strain energy of 0.75 kJ/mol. Moreover, an analysis of the computed 20 nm$^3$ Al polycrystal revealed no correlation between Au segregation energies (and hence the preference for GB sites) and site atomic volume, further suggesting strain relief is not a driving force. Also, given the high homologous temperature used here, we do not expect Au binding to vacancies [80] to play a role, as any vacancies at the GBs will be annealed out and re-distributed among GB sites as free volume.

Thus, all three of these classical driving forces, both individually and together, fail to predict the magnitude of the strong attractive interaction between Au and GB sites in Al in a polycrystal. We stress that the failure does not come from using Miedema's estimates for alloy enthalpies to evaluate $\Delta H_{B\ in\ A}^{seg}$, since our first-principles calculations of the chemical, interfacial, and elastic components, albeit having different magnitudes than Miedema's, provide the same conclusions – the failure is one of missing physics. The quantum-accurate spectral model, on the other hand, aligns with the experiments in both sign and magnitude. However, there are two differences between the models: the classical model is scalar and approximate, while the other model is both quantum-accurate and spectral. It bears separating these two effects in more detail: is spectrality alone enough to explain the experiments, or is quantum accuracy needed? In other words, if spectra of the three above effects (interface energy, chemical interactions, elastic relaxation) were available, would casting the classical model into a spectrum resolve the experimental data?

---

[3] The solid solution enthalpy $\Delta H_{Au\ in\ Al}^{sol+elastic}$ contains both elastic and chemical contributions, $\Delta H_{Au\ in\ Al}^{sol+elastic} = \Delta H_{Au\ in\ Al}^{sol} + \Delta H_{Au\ in\ Al}^{elastic}$. Refer to footnote 2 for details on computing $\Delta H_{Au\ in\ Al}^{sol+elastic}$, including DFT calculation settings, which we maintain throughout the paper. We compute the elastic component from $\Delta H_{Au\ in\ Al}^{elastic} = H(Al_{N-1}Al^{Au}) - N. H_{Al}^{eq}$, where $H(Al_{N-1}Al^{Au})$ is the total unrelaxed energy of the N-atom supercell used to compute the relaxed energy $H(Al_{N-1}Au)$, as detailed in footnote 2, but with the Au atom replaced with an Al atom (without relaxing the supercell); this way, the elastic distortion introduced by a Au atom in an Al supercell is captured without the chemical component. We obtain $\Delta H_{Au\ in\ Al}^{elastic} = 0.75$ kJ/mol. The chemical component is simply $\Delta H_{Au\ in\ Al}^{sol} = \Delta H_{Au\ in\ Al}^{sol+elastic} - \Delta H_{Au\ in\ Al}^{elastic}$, or alternatively, $H(Al_{N-1}Au) - H(Al_{N-1}Al^{Au})$. We obtain $\Delta H_{Au\ in\ Al}^{sol} = -71$ kJ/mol.

We can separate these two effects by considering predictions based on embedded atom (EAM) interatomic potentials. Although such potentials are rarely tuned to predict GB segregation well [18,81], they are generally well suited to capture the above three classical effects; they are tuned to produce surface energies, chemical interactions, elastic constants and atomic sizes in a manner consistent with classical models such as Eqs. (1)-(2). In Fig. 1(a), we show a computed segregation spectrum for Au in Al [29] using the EAM potential from Ref. [64] – one of the few available for Al(Au). This EAM potential is not directly fitted to alloy properties, but instead uses a simplification in which the alloy parameters are assumed to be an average of the elemental ones.

Fig. 1(a) shows the EAM-based GB segregation enthalpy spectrum; like the classical scalar model, it generally suggests that antisegregation is favorable for most sites. Specifically, it predicts that ~84% of GB sites are enthalpically unfavorable to Au segregation, which, for preliminary screening, suggests that Au is at best a weak segregator, with only 16% of sites available for segregation. With a finite concentration and temperature, those few sites compete with entropy, so very little segregation is expected to occur, and indeed antisegregation can prevail. This is illustrated in Fig. 1(b), which shows the calculated solute enrichment using the spectral model, Eq. (4). The EAM-based spectrum predicts a depletion of Au at the GB at 300 °C. Therefore, as expected, the EAM-based spectrum fails to predict the observed experimental segregation of Au, both quantitatively and qualitatively, as shown in Fig. 4.

This illustration underscores a main point of this paper: it is not only the inclusion of a spectrum that is needed to accurately predict GB segregation (although a spectrum is absolutely needed as shown by the range of experimental measurements in a single alloy in Fig. 3). Rather, the calculation of the site segregation enthalpies themselves needs to be as accurate as possible. In the present case quantum accuracy happens to be available, but there are far more spectra available that are based on EAM, which may not be properly tuned to the problem of GB segregation in general. This highlights the need for further development of alloy interatomic potentials that can capture defect energetics with quantum accuracy, as well as the value in scale-bridging computations of defect alloy states. Furthermore, this work underscores the need for further experimental work that probes in depth the variation of solute segregation across the GB space, similar to those in Refs. [5–8,10,82].

Finally, the segregation of Au in Al and its preference for the less coordinated GB sites points to a more complex driving force at the electronic structure level that was only possible to reveal using the quantum-accurate predictions of the spectral approach in Ref. [30]. Al-Au alloys are known to exhibit complex electronic interactions, and thus, one can expect the GBs to amplify this complexity even further. One prime example of such interactions is the $AuAl_2$ intermetallic compound, which has a striking deep purple color [83], and a surprisingly high melting temperature of 1060 °C [66]. To explain the nature of its chemical bonding, Pauling [84] proposed a valence electron charge transfer and compensation mechanism: Au, which is more electronegative than Al, will cause a transfer of non-d-character charge onto it; this is compensated by a back transfer of d-charge from Au to Al to maintain charge neutrality. The end result is a structure with hybridized Al(s, p) and Au(d) valence electrons, and a large depletion of d-electrons in Au atoms [85–87]. While such complex electronic interactions do not explain Al-Au interactions at the GBs, they suggest that a wide range of electronic interactions can take place, including orbital shifts and charge transfer, since the GB network offers a wide variety of bond lengths, neighbor densities, and free volumes for Au atoms in Al. This is an interesting avenue for future research.

## 5. Conclusion

In conclusion, we have shown the limitations of the classical McLean-style approach, as well as the physical motivators traditionally used in the literature, to understand and predict solute segregation in alloys. With the aid of a quantum-accurate treatment of segregation in polycrystals, we were able to uncover and experimentally validate strong GB segregation that is not predicted by conventional wisdom for Au in Al. Our main findings are:

- In dilute Al(Au), Au shows strong enrichment at the GBs with Au concentrations being more than two orders of magnitude higher than the bulk lattice solid solution. The observed concentrations show a strong spectral behavior with a large spread between different GBs, and between different GB locations within the same GB.
- The classical driving forces attributed to solute segregation, namely, the chemical, interfacial, and elastic contributions fail to predict Au segregation in Al. Evaluations of the magnitude of the three contributions using both Miedema's alloy enthalpies and first-principles calculations reach the same conclusion.
- The ab-initio based spectral model provides predictions that are in agreement with the experiments, both qualitatively and quantitatively. In fact it is the only model explored here that is able to predict Au segregation in Al. Its strong predictive capabilities are shown to be the result of being (i) spectral, and (ii) quantum accurate.

Our findings clearly illustrate the value and necessity of advanced computation and learning frameworks to continue building comprehensive databases for GB segregation that provide a robust basis for alloy design and optimization.

## Acknowledgments

This work was supported by the US Department of Energy, Office of Basic Energy Sciences under grant number DE-SC0020180. The microscopy work was performed using MIT.nano Characterization facilities. The authors thank Dr. Yang Yu for his help with the STEM characterization.